\documentclass{elsart}
\usepackage{amssymb}
\usepackage{amsfonts}
\usepackage{amsmath}

\input{tcilatex}

\begin{document}

\begin{frontmatter}



\title{Single particle calculations for a Woods-Saxon potential 
with triaxial deformations, and  large Cartesian oscillator basis}


\author{B. MOHAMMED-AZIZI\thanksref{label1}}
\address{Centre Universitaire de Bechar, Route de Kenadsa, Bechar, 8000,  Algeria}
\thanks[label1]{Corresponding author, E-mail: aziziyoucef@voila.fr}

\author{D.E. MEDJADI\thanksref{label2}}
\address{Ecole Normale Superieure, BP92, Vieux Kouba , Algiers, 16050, Algeria}
\thanks[label2]{E-mail: demedjadi@voila.fr}
\begin{abstract}

 We present a computer program which solves the Schrodinger equation of the stationary states for an average nuclear potential of Woods-Saxon type. In this work, we take specifically into account  triaxial (i.e. ellipsoidal ) nuclear surfaces. The deformation is specified by the usual Bohr parameters.
The calculations are carried out in two stages. In the first, one calculates the representative matrix of the Hamiltonian in the cartesian oscillator basis. In the second stage one diagonalizes this matrix with the help of subroutines of the EISPACK library.
If it is wished, one can calculate all eigenvalues, or only the part of the eigenvalues that are contained in a fixed interval defined in advance. In this latter case the eigenvectors are given conjointly.
The program is very rapid, and the run-time is mainly used for the diagonalization. Thus,  it is possible to use a  significant number of the basis states in  order to insure a best convergence of the results.

\end{abstract}

\end{frontmatter}


{\Large Program summary}{\normalsize \newline
}\newline
{\normalsize Title of program: Triaxial\newline
}\newline
{\normalsize Catalogue number:\newline
}\newline
{\normalsize Licensing provisions: none}\newline
\newline
{\normalsize Computer: PC. AMD Athlon 1000MHz\newline
}\newline
{\normalsize Hard disk: 40 Go\newline
}\newline
{\normalsize Ram: 256 Mo\newline
}\newline
{\normalsize Swap file: 4 Go\newline
}\newline
{\normalsize Operating system: WINDOWS XP\newline
}\newline
{\normalsize Software used: Microsoft Visual FORTRAN 5.0A (with full
optimizations in the settings project options)\newline
}\newline
{\normalsize Programming language: fortran 77/90 (double precision)\newline
}\newline
{\normalsize Number of bits in a word: 32}\newline
\newline
{\normalsize Number of lines: 3150 lines with comments\newline
}\newline
\newline
{\normalsize Nature of the problem:}\newline
{\normalsize The Single particle energies and the single particle wave
functions are calculated from one-body Hamiltonian including a central field
of Woods-Saxon type, a spin-orbit interaction, and the Coulomb potential for
the protons.}\newline
{\normalsize We consider only ellipsoidal (triaxial) shapes. The deformation
of the nuclear shape is fixed by the usual Bohr parameters }$(\beta ,\gamma
) ${\normalsize \ .}\newline
\newline
{\normalsize Method of solution:}\newline
{\normalsize The representative matrix of the Hamiltonian is built by means
of the Cartesian basis of the anisotropic harmonic oscillator, and then
diagonalized by a set of subroutines of the EISPACK library.}\newline
{\normalsize Two quadrature methods of Gauss are employed to calculate
respectively the integrals of the matrix elements of the Hamiltonian, and
the integral defining the Coulomb potential\newline
}\newline
{\normalsize Restrictions:}\newline
{\normalsize There are two restrictions for the code:\newline
The number of the major shells of the basis does not have to exceed Nmax=26.%
\newline
For the largest values of Nmax (}$\sim ${\normalsize 23-26), the
diagonalization takes the major part of the running time, but the global
run-time remains reasonable. \newline
}\newline
{\normalsize Typical running time:\newline
(With full optimization in the project settings of the Microsoft Visual
Fortran 5.0A on Windows XP )\newline
With NMAX=23, for the neutrons case, and for both parities, if we need all
eigenenergies and all eigenfunctions of the bound states, the running time
is about 80 sec on the AMD ATHLON computer at 1GHz. In this case, the
calculation of the matrix elements takes only about 20 sec.\newline
If all unbound states are required, the runtime becomes larger.}

\begin{quote}
{\LARGE Long write-up}
\end{quote}

\section{Purpose of the Fortran program}

\subsection{The Schrodinger equation}

The program solves the Schrodinger equation for one body-deformed potential:%
\begin{equation}
H\left\vert \phi _{i}\right\rangle =E_{i}\left\vert \phi _{i}\right\rangle
\label{1}
\end{equation}%
Here H represents the Hamiltonian of the system (neutrons or protons), and, $%
E_{i}$ , and, $\phi _{i}$ , represent respectively, its eigenenergies, and
its eigenfunctions.

\subsection{The Hamiltonian}

The Hamiltonian of the nucleon is defined by \cite{1,2}:%
\begin{equation}
H=T+V+V^{SO}+e\phi ^{C}  \label{2}
\end{equation}%
The quantities T, V, V$^{\mathit{so}}$, indicate respectively, the kinetic,
potential , and spin-orbit energy. For the proton, the Coulomb potential is
represented by $\phi ^{C}$, and $e$ is its charge\newline
Explicitely:%
\begin{equation}
T=-\frac{\hbar ^{2}}{2m}.\vec{\nabla}^{2}  \label{3}
\end{equation}%
$\hbar $ = Planck constant\newline
m= nucleon masse\newline
Owing to the fact that the nuclear forces have a short-range character, the
average nuclear potential must " follow on average" the nuclear density
distribution:\newline
For the case of the spherical symmetry (see also (\ref{20})) , we have:

\begin{equation}
V(r)\left( \propto \frac{\rho _{_{0}}}{1+\exp (\frac{r-R}{a})}\right) =\frac{%
V_{_{0}}}{1+\exp (\frac{r-R}{a})}
\end{equation}%
For the deformed case, the above definition is generalized as \cite{9}:%
\begin{equation}
V(\vec{r})=\frac{V_{0}}{1+\exp (R_{V}\ L_{V}/a_{V})}  \label{4}
\end{equation}%
with\newline
$V_{0},R_{V},a_{V}$ = mean field parameters\newline
$L_{V}$ = quasi-radius (see eq. (\ref{8}))

\begin{equation}
V^{so}(\vec{r})=-\frac{1}{\hslash }\left( \vec{\nabla}S(\vec{r})\wedge \vec{p%
}\right) \vec{\sigma}  \label{5}
\end{equation}%
with\newline
$\vec{p}=(\hslash /i)\ \vec{\nabla}=$ Neutron or proton momentum\newline
$\overrightarrow{\sigma }=(\sigma _{x},\sigma _{y},\sigma _{z})$ = Pauli
spin-matrices\newline
For the same reasons that for V, the mean field in the expression of the $%
V^{so}$ operator is given by a similar definition (see also the subsection
3.1) :%
\begin{equation}
S(\vec{r})=\frac{\kappa }{1+\exp (R_{so}L_{so}/a_{so})}  \label{6}
\end{equation}%
with\newline
$\kappa $ = spin-orbit coupling strength (there, the quantity $S_{0}$ is
absorbed by $\kappa $, this latter is integrated to $S(\overrightarrow{r})$
and expressed in $MeV.fm^{2}$)\newline
$R_{so},a_{so}$ = mean field parameters of $S(\vec{r})$ \newline
$L_{so}$ is the quasi-radius of the spin-orbit mean field (see eq. (\ref{11}%
))\newline

\section{The Coulomb potential:}

For the protons, the Coulomb's potential is approximated by the one of the
liquid drop model \cite{1,2}%
\begin{multline}
\Phi ^{C}(Z,P,\Phi )=\frac{\rho _{charge}}{4}\int\limits_{Z_{1}}^{Z_{2}}dz%
\times  \label{7} \\
\int\limits_{0}^{2\pi }d\varphi \left[ \frac{-(z-Z)\frac{\partial \rho
_{S}^{2}}{\partial z}+2\rho _{S}^{2}-2\rho _{S}P\cos (\varphi -\Phi )-2P%
\frac{\partial \rho _{S}}{\partial \varphi }\sin (\varphi -\Phi )}{\sqrt{%
(z-Z)^{2}+\rho _{S}^{2}+P^{2}-2\rho _{S}P\cos (\varphi -\Phi )}}\right]
\end{multline}%
where:\newline
$(Z,P,\Phi )$ = cylindrical coordinates of the point where the Coulomb
potential is calculated.(here, $Z$ must not be confused with the protons
number)\newline
$\rho _{charge}$ = $(\acute{Z}-1)e/(4/3)\pi R_{ch}^{3}$= charge density of
the liquid drop\newline
$(\acute{Z}-1)$= "number of protons in the liquid drop"\newline
$R_{ch}^{{}}$= radius of the charge density\newline
The integration domain is defined by the volume limited by the surface $\pi
_{ch}=0$ (see eq. (\ref{14})).\newline
The \textquotedblleft nuclear surface of the protons\textquotedblright\ is
given by $\rho _{S}=\rho _{Surface}$ in the eq. (\ref{15}).\newline
In fact, the code computes directly the quantity $e\Phi ^{C}$ (which is the
Coulomb energy of the proton in the Coulomb field) instead the Coulomb
potential $\Phi ^{C}.$( see the function \emph{ephi}. in the code).

\section{Supplementary details on the deformation of the mean field, and the
different nuclear surfaces:}

\subsection{The quasi-radius and the nuclear surfaces}

In the central average potential, the information on the distortion of the
nuclear surface is given by the dimensionless quasi-radius $L_{V}(\vec{r})$.
which is defined as \cite{1}:%
\begin{equation}
L_{V}(\vec{r})=\frac{\Pi _{V}(\vec{r})}{R_{V}\left\Vert \vec{\nabla}\Pi _{V}(%
\vec{r})\right\Vert }  \label{8}
\end{equation}%
The quantity $\Pi _{V}(\vec{r})$ is defined so that to recover the
well-known spherical case (see section 4).\newline
\begin{equation}
\Pi _{V}(\vec{r})=\sqrt{\pi _{V}(\vec{r})-\pi _{V\min }}-\sqrt{-\pi _{V\min }%
}  \label{9}
\end{equation}%
Here, $\pi _{V\min }$ is the absolute minimum of $\pi _{V}(\vec{r})$. This
latter describes an hypersurface which is not the real nuclear surface,
because generally $\pi _{V}(\overrightarrow{r})\neq 0$ in the expression of
the quasi radius. The actual nuclear surface may be obtained by putting $\pi
_{V}(\overrightarrow{r})=0.$\newline
In this work, we have restricted ourselves only to simple ellipsoidal
(triaxial) shapes for the effective nuclear surface.%
\begin{equation}
\pi _{V}(\vec{r})=\frac{x^{2}}{A_{V}^{2}}+\frac{y^{2}}{B_{V}^{2}}+\frac{z^{2}%
}{C_{V}^{2}}-1=0  \label{10}
\end{equation}%
$A_{V},B_{V},$ and, $C_{V}$ are thus the semi-axes of the ellipsoid.\newline
In fact, we have to consider three distinct interactions, i.e. the central,
the spin-orbit, and the Coulomb interaction. Therefore, we must define three
respective surfaces (equations (\ref{10}),(\ref{13}),(\ref{14})). Thus, the
equation (\ref{10}) describes the nuclear surface relatively to the central
interaction $V(\overrightarrow{r})$.\newline
In completely analogous way, we have to define similar quantities for the
spin-orbit interaction 
\begin{equation}
L_{so}(\vec{r})=\frac{\Pi _{so}(\vec{r})}{R_{so}\left\Vert \vec{\nabla}\Pi
_{so}(\vec{r})\right\Vert }  \label{11}
\end{equation}%
with,%
\begin{equation}
\Pi _{so}(\vec{r})=\sqrt{\pi _{so}(\vec{r})-\pi _{so\min }}-\sqrt{-\pi
_{so\min }}  \label{12}
\end{equation}%
Where, $\pi _{so\min }$ is the absolute minimum of $\pi _{so}(\vec{r})$, and
the effective "spin-orbit surface" is written as:%
\begin{equation}
\pi _{so}(\vec{r})=\frac{x^{2}}{A_{so}^{2}}+\frac{y^{2}}{B_{so}^{2}}+\frac{%
z^{2}}{C_{so}^{2}}-1=0  \label{13}
\end{equation}%
For the Coulomb potential, the effective nuclear surface is defined \ in the
same way:%
\begin{equation}
\pi _{ch}(\vec{r})=\frac{x^{2}}{A_{ch}^{2}}+\frac{y^{2}}{B_{ch}^{2}}+\frac{%
z^{2}}{C_{ch}^{2}}-1=0  \label{14}
\end{equation}%
Following the expression \ of the equation (\ref{7}), the Coulomb potential
must be expressed in cylindrical coordinates Therefore, the equation of the
effective "Coulomb nuclear surface" (\ref{14}) can be rewritten as:%
\begin{equation}
\rho _{surface}^{2}=\frac{1-\left( z/C_{ch}\right) ^{2}}{\left\{ \left( \cos
\varphi /A_{ch}\right) ^{2}+\left( \sin \varphi /B_{ch}\right) ^{2}\right\} }
\label{15}
\end{equation}%
where:

$x=\rho _{surface}\cos \varphi $ , $\ y=\rho _{surface}\sin \varphi $ , $z=z$%
\newline
The "three densities"(neutrons, protons, spin-orbit) differ very little from
each other. Therefore, the three surfaces are homothetic and slightly
different to each other. Nevertheless, in order to simplify the problem, the
protons distribution is assumed to be uniform in the calculation of the
Coulomb potential.

\subsection{The deformation parameters}

Since we have three similar surfaces, and, so as to avoid repeating three
times the same thing, we will omit to specify the indices of the surfaces.
For example, the three volume conservation conditions are simply replaced by
only one equation :%
\begin{equation}
\frac{4}{3}\pi ABC=\frac{4}{3}\pi R^{3}  \label{16}
\end{equation}%
Actually, because of this condition, only two parameters are necessary%
\newline
As usual, one will prefer the Bohr parameters $\left( \beta ,\gamma \right) $
instead of those of the ellipsoid. 
\begin{subequations}
\begin{align}
A& =\frac{R}{\chi }\left[ 1+\beta \left( \frac{5}{4\pi }\right) ^{1/2}\cos
(\gamma -\frac{2}{3}\pi )\right]  \label{17} \\
B& =\frac{R}{\chi }\left[ 1+\beta \left( \frac{5}{4\pi }\right) ^{1/2}\cos
(\gamma -\frac{4}{3}\pi )\right]  \label{18} \\
C& =\frac{R}{\chi }\left[ 1+\beta \left( \frac{5}{4\pi }\right) ^{1/2}\cos
(\gamma )\right]  \label{19}
\end{align}

$R$ is the radius, and, $\chi $ insures the volume conservation condition (%
\ref{16}),

\section{The Spherical case. The mean field parameters}

\subsection{The case of spherical symmetry}

When A=B=C , or when the Bohr parameter $\beta =0$, the nuclear surface
becomes spherical, and $R_{V}$, $R_{so}$, or, $R_{ch}$ represents simply the
nucleus radius. In this case, we obtain the familiar Fermi-function for the
two mean potentials ((\ref{4}),(\ref{6})): 
\end{subequations}
\begin{equation}
V(\overrightarrow{r})=\frac{V_{0}}{1+\exp \left[ \left( r-R_{V}\right) /a_{V}%
\right] }\qquad S(\vec{r})=\frac{\kappa }{1+\exp \left[ \left(
r-R_{so}\right) /a_{so}\right] }  \label{20}
\end{equation}%
i.e. potentials of Woods-Saxon type. \newline
The Coulomb's potential (\ref{7}) reduces to the well-known form:%
\begin{equation}
\begin{tabular}{l}
$\Phi ^{c}(r)=\left[ (Z-1)e/2R_{ch}\right] \left[ 3-(r/R_{ch})^{2}\right] $%
\qquad if$\qquad r\leq R_{ch}$ \\ 
$\Phi ^{c}(r)=(Z-1)e/r$\qquad if\qquad $r\geq R_{ch}$%
\end{tabular}
\label{21}
\end{equation}%
\newline
The spin-orbit interaction (\ref{5}) can be expressed in the spherical case
as:\newline
\begin{equation}
V^{so}=-(\frac{\partial S(r)}{\partial r}\frac{\overrightarrow{r}}{r}\wedge 
\overrightarrow{p})\overrightarrow{\sigma }=-\frac{1}{r}\frac{\partial S(r)}{%
\partial r}(\overrightarrow{r}\wedge \overrightarrow{p})\overrightarrow{%
\sigma }
\end{equation}%
\newline
Finally, the $V^{so}$ operator takes the familiar form,%
\begin{equation}
V^{so}(\overrightarrow{r})=-\frac{1}{r}\frac{\partial S(r)}{\partial r}\vec{l%
}.\vec{\sigma}  \label{22}
\end{equation}

The relations (\ref{20}-\ref{22}) of the spherical case are not used
explicitly in the code. However, the well known spherical degeneracy of the
energy levels were used in order to check the program. Moreover, the
relation (\ref{21}) serves as a first checking for the Coulomb potential.

\subsection{Mean field parameters}

Two options have been included in the code in order to choose between a
particular set of parameters, or the Myers parameters \cite{3}. Thus, it is
possible to define its own parameters in a separate file, or to employ those
of Myers. In this latter case, the calculations are made in a suitable
subroutine. In fact, only a part of the parameters set, namely, $%
V_{0},R_{V},R_{so},R_{ch}$ , is deduced from the droplet model of Myers, the
remaining, i.e. $\kappa ,a_{V\text{,}}a_{so}$, are extracted from the Ref. 
\cite{1}. The explicit expressions for these parameters are given in
appendix.

\section{Principle of resolution}

\subsection{The method}

The principle of this method consists to look for the eigenfunctions of the
Schrodinger equation by their expansion on a truncated basis of the harmonic
oscillator. In other words, the method used in solving such problem amounts
essentially to writing the representative matrix of the Hamiltonian in this
basis.\newline
In practice, this method is characterized by two distinct stages. First, one
builds the representative matrix of the Hamiltonian by means of the cited
basis. Next, one diagonalizes this matrix in order to obtain the eigenvalues
and the eigenvectors.

In our work, the cartesian coordinates are the most suitable.

\subsection{The harmonic oscillator basis}

The basis functions of the harmonic oscillator are defined as:%
\begin{equation}
\left\vert n_{x}n_{y}n_{z}\Sigma \right\rangle \equiv i^{ny}\phi
_{n_{x}}(x).\phi _{n_{y}}(y).\phi _{n_{z}}(z).\vec{\sigma}_{\Sigma }
\label{23}
\end{equation}%
There, $i^{n_{y}}$ is a phase factor which insures, in accordance with the
imposed symmetries (see section 6), that the matrix elements are real.%
\newline
Explicitly:

\begin{equation}
\phi _{n_{x}}(x)=\sqrt{\beta _{x}}\exp \left[ -\left( \beta _{x}x\right)
^{2}/2\right] .h_{n_{x}}(\beta _{x}x)\quad with\quad \beta _{x}=\sqrt{\frac{%
m\omega _{x}}{\hbar }}  \label{24}
\end{equation}%
with analogous expressions for the y, and z axes.\newline
The intrinsic spin states are:%
\begin{equation}
\vec{\sigma}_{+1/2}=\left[ 
\begin{array}{c}
1 \\ 
0%
\end{array}%
\right] ,\qquad \vec{\sigma}_{-1/2}=\left[ 
\begin{array}{c}
0 \\ 
1%
\end{array}%
\right]  \label{25}
\end{equation}%
The $h_{n_{x}}$, (or $h_{n_{y}},$ or $h_{n_{z}}$) quantities symbolize the
normalized Hermite polynomials.

\begin{equation}
h_{n_{x}}(x)=H_{n_{x}}(x)/\sqrt{\left( 2^{n_{x}}.n_{x}!\pi ^{1/2}\right) }
\label{26}
\end{equation}%
$H_{n_{x}}(x)$ are the usual Hermite polynomials\newline
The quantum numbers $n_{x}$ , $n_{y}$ , $n_{z}$ , are integers, and give the
order of the Hermite polynomials, $\Sigma =\pm \frac{1}{2}$ represents the
projection of the intrinsic spin on the z-axis.\newline
At last, $m$ and $\omega _{x}$ , $\omega _{y}$ , $\omega _{z}$ represents
the mass of the oscillator, i.e. the mass of the nucleon, and, its
frequencies.\newline
For convenient, the quantities $\hbar \omega _{x}$, $\hbar \omega _{y}$, $%
\hbar \omega _{z}$,(or respectively $\beta _{x}$,$\beta _{y}$,$\beta _{z}$)
are called the deformation parameters of the basis.\newline
If the three frequencies are equal, the oscillator is then isotropic, and it
can be characterized by only one frequency ($\omega _{x}=\omega _{y}=\omega
_{z}=\omega _{0}$)\newline
Furthermore, we assume that the nuclear surface of the oscillator is an
equipotential. This involves the following condition:%
\begin{equation}
\omega _{x}\omega _{y}\omega _{z}=\omega _{0}^{3}\qquad or\qquad \left(
\hbar \omega _{x}\right) .\left( \hbar \omega _{y}\right) .\left( \hbar
\omega _{z}\right) =\left( \hbar \omega _{0}\right) ^{3}  \label{27}
\end{equation}

\subsection{The representative matrix of the Hamiltonian}

With help of this basis, the matrix elements of the Hamiltonian $H$ can be
written as:%
\begin{equation}
\left\langle n_{x}^{\prime }n_{y}^{\prime }n_{z}^{\prime }\Sigma ^{\prime
}\right\vert H\left\vert n_{x}n_{y}n_{z}\Sigma \right\rangle =\left\langle
n_{x}^{\prime }n_{y}^{\prime }n_{z}^{\prime }\Sigma ^{\prime }\right\vert
T+V+V^{so}+e.\Phi ^{c}\left\vert n_{x}n_{y}n_{z}\Sigma \right\rangle
\label{28}
\end{equation}

\subsubsection{Matrix elements of the mean field V and the Coulomb energy e.$%
\Phi ^{C}$}

Since $V$ does not depend on the spin, we adopt the following convenient
notation for V:%
\begin{equation}
\left\langle n_{x}^{\prime }n_{y}^{\prime }n_{z}^{\prime }\Sigma ^{\prime
}\right\vert V\left\vert n_{x}n_{y}n_{z}\Sigma \right\rangle =\left(
n_{x}^{\prime }n_{y}^{\prime }n_{z}^{\prime }\left\vert V\right\vert
n_{x}n_{y}n_{z}\right) \times \delta _{\Sigma ^{\prime }\Sigma }  \label{29}
\end{equation}

where:%
\begin{multline}
\left( n_{x}^{\prime }n_{y}^{\prime }n_{z}^{\prime }\left\vert V\right\vert
n_{x}n_{y}n_{z}\right)  \label{30} \\
=i^{(n_{y}-n_{y}^{\prime })}\iiint \phi _{n_{x}^{\prime }}(x)\phi
_{n_{y}^{\prime }}(y)\phi _{n_{z}^{\prime }}(z)V(x,y,z)\phi _{n_{x}}(x)\phi
_{n_{y}}(y)\phi _{n_{z}}(z)dxdydz= \\
=i^{(n_{y}-n_{y}^{\prime })}\int\limits_{-\infty }^{+\infty
}\int\limits_{-\infty }^{+\infty }\int\limits_{-\infty }^{+\infty
}e^{-(x^{2}+y^{2}+z^{2})}h_{n_{x}^{^{\prime }}}(x).h_{n_{y}^{^{\prime
}}}(y).h_{n_{z}^{^{\prime }}}(z) \\
\times V(\frac{x}{\beta _{x}},\frac{y}{\beta _{y}},\frac{z}{\beta _{z}}%
).h_{n_{x}}(x).h_{n_{y}}(y).h_{n_{z}}(z).dxdydz
\end{multline}%
\newline
\newline
with $\ \beta _{x}$, $\beta _{y}$, $\beta _{z}$defined in eq.(\ref{24})

\begin{itemize}
\item Due to the parity of $V$ , the integral (\ref{30}) vanishes if one of
the three following conditions is not fulfilled:\newline
$(-1)^{n_{x}}=(-1)^{n_{x}^{\prime }}$\newline
$(-1)^{n_{y}}=(-1)^{n_{y}^{\prime }}$\newline
$(-1)^{n_{z}}=(-1)^{n_{z}^{\prime }}$\newline
Therefore, it is not necessary to calculate them.\newline
Since the respective indices must have the same parity, the complex factor
of (\ref{30}) can be rewritten as:\newline
$i^{(n_{y}^{\prime }-n_{y})}=(-1)^{(n_{y}^{\prime }-n_{y})/2}$

\item Although the $\left( n_{x}^{\prime }n_{y}^{\prime }n_{z}^{\prime
}\left\vert V\right\vert n_{x}n_{y}n_{z}\right) $ elements are spin
independent, they are stored actually as:$\left\langle n_{x}^{\prime
}n_{y}^{\prime }n_{z}^{\prime }\Sigma ^{\prime }\right\vert V\left\vert
n_{x}n_{y}n_{z}\Sigma \right\rangle $ in the computer memory

\item The matrix elements of the Coulomb energy are calculated in the same
way as those of $V(\overrightarrow{r})$ (putting $e\phi ^{c}(\overrightarrow{%
r})$ instead $V(\overrightarrow{r})$), with the same change of scale.

\item In the gaussian integration, we have to calculate the Hermite
polynomials, the central mean potential $V(\overrightarrow{r})$, the
spin-orbit mean potential $S(\overrightarrow{r})$, and the Coulomb potential 
$\phi ^{c}(\overrightarrow{r})$ only at the nodes. Therefore, it is more
convenient to store these quantities in specific arrays before any
calculations (see the common/tabh/... declaration in the subroutine setsub).
\end{itemize}

\subsubsection{Matrix elements of the spin-orbit energy operator V$^{so}$}

Due to the presence of the derivative of $S(\overrightarrow{r})$, the direct
calculation of the matrix elements of $V^{so}$ , i.e. $\left\langle
n_{x}^{\prime }n_{y}^{\prime }n_{z}^{\prime }\Sigma ^{\prime }\right\vert
V^{so}\left\vert n_{x}n_{y}n_{z}\Sigma \right\rangle $, is not convenient.
These derivatives can be transferred on the basis functions by partial
integration. Therefore, the derivatives of the basis functions can be
expressed from the basis functions themselves by mean of the recursion
relations. Finally, the matrix elements of $V^{so}$ can be obtained by
suitable combinations of $\left( n_{x}^{\prime }n_{y}^{\prime }n_{z}^{\prime
}\left\vert S\right\vert n_{x}n_{y}n_{z}\right) $ elements, i.e.%
\begin{multline}
\left\langle n_{x}^{\prime }n_{y}^{\prime }n_{z}^{\prime }\Sigma ^{\prime
}\left\vert V^{so}\right\vert n_{x}n_{y}n_{z}\Sigma \right\rangle =
\label{35} \\
\frac{m\omega _{0}}{2\hbar }\left[ 2B_{z}\left( \Sigma ^{\prime }\left\vert
\sigma _{z}\right\vert \Sigma \right) +B_{+}\left( \Sigma ^{\prime
}\left\vert \sigma _{-}\right\vert \Sigma \right) +B_{-}\left( \Sigma
^{\prime }\left\vert \sigma _{+}\right\vert \Sigma \right) \right]
\end{multline}%
where:%
\begin{equation}
\sigma _{\pm }=\sigma _{x}\pm \sigma _{y}  \label{36}
\end{equation}%
\begin{equation}
B_{\pm }=B_{x}\mp B_{y}  \label{37}
\end{equation}%
\newline
\begin{multline}
B_{x}=\sqrt{\frac{\omega _{y}\omega _{z}}{\omega _{0}^{2}}}[-\sqrt{%
n_{y}^{\prime }(n_{z}+1)}\left( n_{x}^{\prime }n_{y}^{\prime
}-1,n_{z}^{\prime }\left\vert S\right\vert n_{x}n_{y}n_{z}+1\right)
\label{38} \\
-\sqrt{n_{y}(n_{z}^{\prime }+1)}\left( n_{x}^{\prime }n_{y}^{\prime
}n_{z}^{\prime }+1\left\vert S\right\vert n_{x}n_{y}-1,n_{z}\right) \\
+\sqrt{n_{z}(n_{y}^{\prime }+1)}\left( n_{x}^{\prime }n_{y}^{\prime
}+1,n_{z}^{\prime }\left\vert S\right\vert n_{x}n_{y}n_{z}-1\right) \\
+\sqrt{n_{z}^{\prime }(n_{y}+1)}\left( n_{x}^{\prime }n_{y}^{\prime
},n_{z}^{\prime }-1\left\vert S\right\vert n_{x}n_{y}+1,n_{z}\right) ]%
\newline
\end{multline}%
\newline
\begin{multline}
B_{y}=\sqrt{\frac{\omega _{z}\omega _{x}}{\omega _{0}^{2}}}[-\sqrt{%
n_{z}^{\prime }(n_{x}+1)}\left( n_{x}^{\prime },n_{y}^{\prime }n_{z}^{\prime
}-1\left\vert S\right\vert n_{x}+1,n_{y}n_{z}\right)  \label{39} \\
+\sqrt{n_{z}(n_{x}^{\prime }+1)}\left( n_{x}^{\prime }+1,n_{y}^{\prime
}n_{z}^{\prime }\left\vert S\right\vert n_{x}n_{y}n_{z}-1\right) \\
-\sqrt{n_{x}(n_{z}^{\prime }+1)}\left( n_{x}^{\prime }n_{y}^{\prime
}n_{z}^{\prime }+1\left\vert S\right\vert n_{x}-1,n_{y},n_{z}\right) \\
+\sqrt{n_{x}^{\prime }(n_{z}+1)}\left( n_{x}^{\prime }-1,n_{y}^{\prime
}n_{z}^{\prime }\left\vert S\right\vert n_{x}n_{y}n_{z}+1\right) ]\newline
\end{multline}%
\newline
\begin{multline}
B_{z}=\sqrt{\frac{\omega _{x}\omega _{y}}{\omega _{0}^{2}}}[-\sqrt{%
n_{x}^{\prime }(n_{y}+1)}\left( n_{x}^{\prime }-1,n_{y}^{\prime
}n_{z}^{\prime }\left\vert S\right\vert n_{x}n_{y}+1,n_{z}\right)  \label{40}
\\
-\sqrt{n_{x}(n_{y}^{\prime }+1)}\left( n_{x}^{\prime }n_{y}^{\prime
}+1n_{z}^{\prime }\left\vert S\right\vert n_{x}-1,n_{y}n_{z}\right) \\
+\sqrt{n_{y}(n_{x}^{\prime }+1)}\left( n_{x}^{\prime }+1,n_{y}^{\prime
}n_{z}^{\prime }\left\vert S\right\vert n_{x}n_{y}-1,n_{x}\right) \\
+\sqrt{n_{y}^{\prime }(n_{x}+1)}\left( n_{x}^{\prime }n_{y}^{\prime
}-1,n_{x}^{\prime }\left\vert S\right\vert n_{x}+1,n_{y}n_{x}\right) \newline
\end{multline}%
where $S(\overrightarrow{r})$ is given by eq.(\ref{6})

\begin{itemize}
\item The changes of sign in $B_{y}$ are involved by the phase factor of the
basis functions.

\item The computations of the matrix elements of $S(\overrightarrow{r})$ are
carried out like those of $V(\overrightarrow{r})$) in (\ref{30}).
\end{itemize}

\subsubsection{\protect\bigskip Matrix elements of the kinetic energy
operator T}

At last, the matrix elements of the $T$ operator can be calculated in
straightforward way:

\begin{align}
\left\langle n_{x}^{\prime }n_{y}^{\prime }n_{z}^{\prime }\Sigma ^{\prime
}\left\vert T\right\vert n_{x}n_{y}n_{z}\Sigma \right\rangle & =\frac{1}{4}%
\delta _{\Sigma \Sigma ^{\prime }}i^{(n_{y}-n_{y}^{\prime })}\times  \notag
\\
& [\hslash \omega _{z}\delta _{n_{x}n_{x}^{\prime }}\delta
_{n_{y}n_{y}^{\prime }}\delta _{n_{z}n_{z}^{\prime }}\text{ }(2n_{z}+1) 
\notag \\
& -\hslash \omega _{z}\delta _{n_{x}n_{x}^{\prime }}\delta
_{n_{y}n_{y}^{\prime }}\delta _{n_{z}n_{z}^{\prime }+2}\text{ }\sqrt{%
n_{z}^{\prime }(n_{z}+1)}  \notag \\
& -\hslash \omega _{z}\delta _{n_{x}n_{x}^{\prime }}\delta
_{n_{y}n_{y}^{\prime }}\delta _{n_{z}n_{z}^{\prime }-2}\text{ }\sqrt{%
n_{z}(n_{z}^{\prime }+1)}  \notag \\
& +\ cyclic\ permutations]  \label{41}
\end{align}

\section{Symmetry properties of the nuclear surface}

The three surfaces can be written as:%
\begin{equation}
\pi =\frac{x^{2}}{A^{2}}+\frac{y^{2}}{B^{2}}+\frac{z^{2}}{C^{2}}-1=0
\label{42}
\end{equation}%
This implies the following properties:%
\begin{equation}
\pi (x,y,z)=\pi (-x,y,z)=\pi (x,-y,z)=\pi (x,y,-z)  \label{43}
\end{equation}%
Thus, for the two mean fields (i.e. central and spin-orbit fields), and for
the Coulomb potential, we obtain:\newline
\begin{equation}
V(-x,-y,-z)=V(x,y,z)  \label{100}
\end{equation}%
\newline
\begin{equation}
S(-x,-y,-z)=S(x,y,z)  \label{101}
\end{equation}

\begin{equation}
\Phi ^{C}(-x,-y,-z)=\Phi ^{C}(x,y,z)  \label{102}
\end{equation}

\subsection{Parity}

Because of the relations (\ref{100}), (\ref{101}), and (\ref{102}) the
parity is a good quantum number, and the initial matrix decays into two
sub-matrices according to the number%
\begin{equation}
p_{a}=(-1)^{n_{x}+n_{y}+n_{z}}=\pm 1  \label{44}
\end{equation}%
\newline
Obviously, if $(n_{x}+n_{y}+n_{z})$ is even or odd the parity is
respectively positive or negative.

\subsection{Signature}

Furthermore, the Kramers degeneracy is expressed here, by the fact that the
eigenvalues are doubly degenerated relatively to the signature quantum
number $q_{K}$ , which is defined by:%
\begin{equation}
q_{K}=(-1)^{n_{x}+n_{y}}\Sigma =\pm 1/2  \label{45}
\end{equation}%
Consequently, the secular matrix splits into two sub-matrices, and only one
must be considered. The two matrices contain the same set of eigenvalues,
but the eigenfunctions are time-reversed each other.

\subsection{Consequences of these symmetries}

The computer code carries out calculations only for one kind of particles.
Therefore, in order to take into account both neutrons and protons, the code
must be run twice.

Since the Hamiltonian connects only states with the same parity, the
computer code is built in such way that it separates the two types of parity 
$p_{a}=\pm 1$ and performs the calculations separately for them.
Consequently, the representative matrix of the Hamiltonian splits into two
blocks with a definite parity for each block. The diagonalization is then
carried out in each block. \newline
Furthermore, the Kramers degeneracy involves the same eigenvalues for states
which are time-reversed each other. For each block of a definite parity, the
eigenenergies can be separated into two sets defined by the signature $%
q_{K}=\pm 1/2$. The code will make calculations only for $q_{K}=+1/2$. The
second block $q_{K}=-1/2$ will be implicit, and will contain same energies
but with time-reversed eigenfunctions. These eigenvectors may be obtained by
application of the time reversal operator, i.e. by the operator $T=-i\sigma
_{y}K_{0}$ , where $\sigma _{y}$ is a Pauli matrix and $K_{0}$ is the
operator of complex conjugation.\newline
Thus, one obtains 8 blocks, of which 4 are actually calculated (i.e. here
the four first).\newline
\begin{tabular}{cccc}
$1)$ & $\left[ n\right] $ & $\left[ p_{a}=+1\right] $ & $\left[ q_{K}=+1/2%
\right] $%
\end{tabular}%
\newline
\begin{tabular}{cccc}
$2)$ & $\left[ n\right] $ & $\left[ p_{a}=-1\right] $ & $\left[ q_{K}=+1/2%
\right] $%
\end{tabular}%
\newline
\begin{tabular}{cccc}
$3)$ & $\left[ p\right] $ & $\left[ p_{a}=+1\right] $ & $\left[ q_{K}=+1/2%
\right] $%
\end{tabular}%
\newline
\begin{tabular}{cccc}
$4)$ & $\left[ p\right] $ & $\left[ p_{a}=-1\right] $ & $\left[ q_{K}=+1/2%
\right] $%
\end{tabular}%
\newline
\begin{tabular}{cccc}
$5)$ & $\left[ n\right] $ & $\left[ p_{a}=+1\right] $ & $\left[ q_{K}=-1/2%
\right] $%
\end{tabular}%
\newline
\begin{tabular}{cccc}
$6)$ & $\left[ n\right] $ & $\left[ p_{a}=-1\right] $ & $\left[ q_{K}=-1/2%
\right] $%
\end{tabular}%
\newline
\begin{tabular}{cccc}
$7)$ & $\left[ p\right] $ & $\left[ p_{a}=+1\right] $ & $\left[ q_{K}=-1/2%
\right] $%
\end{tabular}%
\newline
\begin{tabular}{cccc}
$8)$ & $\left[ p\right] $ & $\left[ p_{a}=-1\right] $ & $\left[ q_{K}=-1/2%
\right] $%
\end{tabular}%
\newline
So, it is important to point out that, the number of the basis states
practically taken into account by the code is the half of the actual number.

\section{Numerical choices and prescriptions}

\subsection{The quadratures}

The matrix elements of $V(\overrightarrow{r})$, $e\Phi ^{C}(\overrightarrow{r%
})$ and $V^{so}(\overrightarrow{r})$ are calculated with the Gauss-Hermite
method, with $30\times 30\times 30$ of mesh points. The Coulomb potential $%
\Phi ^{C}(\overrightarrow{r})$ is also evaluated numerically by the
Gauss-Legendre method, but with $48\times 48$ of mesh points .

These choices seem to be sufficient relatively to the size of the basis ( $%
N_{\max }\leqq 26$ ), and the interval of deformation ( $0\leq \beta \leq
0.6 $). A direct checking has been done by increasing the number of
quadrature points and by comparing the stability of the results (even with
20 points the results remain very correct).

\subsection{Prescription of the basis truncation}

In practice, the Hamiltonian matrix is finite. Therefore, for reasons of
accuracy, we have to select a sufficient number of the basis states.
Generally, we adopt one of the two following criteria:\newline
The first (spherical criterion) consists in choosing all basis states which
satisfy the following inequality.%
\begin{equation}
n_{x}+n_{y}+n_{z}\leq N_{\max }  \label{46}
\end{equation}%
With this criterion the total number of the basis states is given by
(Nmax+1) x (Nmax+2) x (Nmax+3)/6.\newline
In the second criterion (deformed criterion), one selects the states
according to the deformation of the basis, i.e. according to the three
frequencies of basis.%
\begin{equation}
(n_{x}+\frac{1}{2})\hbar \omega _{x}+(n_{y}+\frac{1}{2})\hbar \omega
_{y}+(n_{z}+\frac{1}{2})\hbar \omega _{z}\leq E_{cut}=(N_{\max }+\frac{3}{2}%
)\hbar \omega _{0}  \label{47}
\end{equation}%
(In fact, these three frequencies are already connected by the condition (%
\ref{27})).\newline
Thus, the choice of $N_{\max }$ determines the size of the basis. The files
"conver12.res" and "conver13.res" give some details about this.

\subsection{Optimization of the basis frequencies}

Since the Hamiltonian operator does not depend on the oscillator
frequencies, its eigenfunctions, and its eigenenergies, must not depend on
these parameters. In practice, the representative matrix of the Hamiltonian
is built by means of a finite number of oscillator eigenfunctions. This
implies a \emph{spurious} dependence according to these parameters.\newline
In another point of view, we might consider this method as a variational
method in which the variational parameters are the frequencies of the basis.
Thus, the best set (in terms of energy) for these frequencies should be
precisely the one, \emph{which minimizes the eigenenergies}, or simply their
sum.\newline
For practical reasons, this method is not easy, since the variation is
three-dimensional. However, it can be often more efficient to use some
prescriptions in order to find (in an economical way) suitable values for
these parameters. \newline
In the present work, we have adopted the approach of the references \cite{1}
and \cite{9}. In that method, we define first, the quantities $p$ and $q$ by:%
\begin{equation}
q^{2}=\frac{\left\langle z^{2}\right\rangle }{\left\langle
x^{2}\right\rangle }=\frac{\int d\tau .\rho (\vec{r})z^{2}}{\int d\tau .\rho
(\vec{r})x^{2}}\qquad \qquad \qquad p^{2}=\frac{\left\langle
z^{2}\right\rangle }{\left\langle y^{2}\right\rangle }=\frac{\int d\tau
.\rho (\vec{r})z^{2}}{\int d\tau .\rho (\vec{r})y^{2}}  \label{48}
\end{equation}%
where $\rho (\vec{r})$ is the nuclear density. Note that the present
definition of $p$ differs from that of the ref.\cite{1}. \newline
For a harmonic oscillator, the equations (\ref{48}) are reduced to very
simple relations.%
\begin{equation}
q_{HO}=\frac{\omega _{x}}{\omega _{z}}\qquad \qquad p_{HO}=\frac{\omega _{y}%
}{\omega _{z}}  \label{49}
\end{equation}%
Next, we have to add, to these two formulas, the relation (\ref{27}). Now,
it is possible to replace the parameter set ( $\omega _{x},\omega
_{y},\omega _{z}$ ) by the equivalent ($q_{HO},p_{HO},\omega _{0}$).\newline
In the same way, for the potential of Woods-Saxon, the nuclear density can
be approximated by the one of the liquid drop (i.e. a constant density). We
obtain thus:%
\begin{equation}
q_{WS}=\frac{c}{a}\qquad \qquad p_{WS}=\frac{c}{b}  \label{50}
\end{equation}%
At last, we \textquotedblleft adapt\textquotedblright\ the oscillator basis
to the nuclear shape by requiring:%
\begin{equation}
q_{HO}=q_{WS}\qquad \qquad p_{HO}=p_{WS}  \label{51}
\end{equation}%
For the $\omega _{0}$ value, we can adopt simply the one of the Nilsson
model.%
\begin{equation}
\hbar \omega _{0}\approx 41.A^{-.\frac{1}{3}}  \label{52}
\end{equation}%
Many tests have shown that relations (\ref{51}) and (\ref{52}) give \emph{%
automatically} very close values to those that produce the "true"
minimization. Furthermore, a general rule is that a large basis size
involves always a weak dependence of the eigenvalues according to these
parameters. Going to the limit, we can say that if the basis was infinite,
the results would be independent to the basis parameters. Conversely, for a
too small basis, the dependance is strong, and the results become too
inaccurate.

For a square well, or (approximately) a Woods-Saxon potential, simple
analytical considerations lead to a more "refined" value for the parameter $%
\hbar \omega _{0}$:%
\begin{equation}
\hbar \omega _{0}\approx \frac{5}{3}\left( \frac{2}{\pi ^{2}}\right)
^{1/3}\left\vert V_{0}\right\vert .A^{-1/3}\approx 0.979\ \left\vert
V_{0}\right\vert .A^{-1/3}  \label{53}
\end{equation}%
where $V_{0}$ is the depth of the potential. The equation (\ref{53}) is
obtained by requiring the condition 
\begin{equation*}
\left\langle r^{2}\right\rangle _{harm.\text{ }Oscillator}=\left\langle
r^{2}\right\rangle _{square\text{ }well}
\end{equation*}%
The averages are made with the semi-classical Thomas-Fermi density:%
\begin{equation*}
\rho _{TF}(r)=\frac{(2m)^{3/2}}{3\pi ^{2}\hbar ^{3}}(\lambda -V(r))^{3/2}
\end{equation*}%
The Fermi level $\lambda $ is determined by the condition of conservation of
the particle number $A$.\newline
The relation (\ref{53}) could explain the empirical scale factor used
sometimes \cite{1,2,5} in the " standard equation" (\ref{52}).

\subsection{The numerical values of the $\protect\beta $ parameters of the
basis}

The quantities $\beta _{0}=\sqrt{\frac{m\omega _{0}}{\hbar }}$ , $\beta _{x}=%
\sqrt{\frac{m\omega _{x}}{\hbar }}$ , $\beta _{y}=\sqrt{\frac{m\omega _{y}}{%
\hbar }}$ , and $\beta _{z}=\sqrt{\frac{m\omega _{z}}{\hbar }}$. are
numerically calculated like 
\begin{equation*}
\sqrt{\frac{m\omega }{\hbar }}=\sqrt{\frac{mc^{2}}{\hbar ^{2}c^{2}}\hbar
\omega }
\end{equation*}%
\newline
The values are:\newline
$m_{p}c^{2}=938.2592\ MeV$\newline
$m_{n}c^{2}=939.553\ MeV$\newline
$\hbar c=197.32879\ MeV\ fm$\newline
This involves:%
\begin{equation}
\frac{m_{p}c^{2}}{(\hbar c)^{2}}=t_{p}=0.0240958315\ MeV^{-1}fm^{-2}
\end{equation}%
\newline
\begin{equation}
\frac{m_{n}c^{2}}{(\hbar c)^{2}}=t_{n}=0.0241290571\ MeV^{-1}fm^{-2}
\end{equation}%
\newline
so that :%
\begin{equation}
\beta _{0}=\sqrt{t.\hbar \omega _{0}},\qquad \beta _{x}=\sqrt{t.\hbar \omega
_{x}},\qquad \beta _{y}=\sqrt{t.\hbar \omega _{y}},\qquad \beta _{z}=\sqrt{%
t.\hbar \omega _{z}}
\end{equation}%
with $t=t_{p}$ or $t=t_{n}$ \newline
The numbers $t_{p}$ and $t_{n}$ appear in the subroutine \textquotedblleft
Basisparam\textquotedblright .

\section{Diagonalization}

The diagonalization of the representative matrix of the Hamiltonian is
carried out by a set of subroutines extracted from the EISPACK library of
FORTRAN programs (http//www.netlib.org/eispack/). Thus, four subroutines of
this library were gathered: \newline
The subroutine tred1 transforms any full symmetrical matrix into a
tridiagonal symmetrical matrix by using the Givens-Householder's method.%
\newline
For a tridiagonal symmetrical matrix the subroutine tql1 uses the ql method
to calculate only the eigenvalues of a tridiagonal matrix. \newline
For a tridiagonal symmetrical matrix, the subroutine tsturm calculates the
eigenvalues contained in a given interval. This subroutine calculates also
the eigenvectors associated to the found eigenvalues. The adopted method is
that of the bisection and the inverse iteration. \newline
Lastly, the subroutine trbak1 recalculates the eigenvectors found by tsturm
relatively to the initial basis (that of tred1). The sought eigenvectors are
thus obtained. \newline
These subroutines are called by the subroutines diagoplus (for the positive
parity), and diagominus (for the negative parity) in which the options of
the diagonalization are specified. These options are indicated in the
comments of the program, and below, in the subsection 10.1.1.

\section{The subroutines and the functions of the Program.}

The program is composed by a main program, 29 subroutines and 6 functions.
The role reserved for each program is briefly described in the paragraph
below (and described again in details in the comments of the program).%
\newline
In fact, all calculations are governed by the subroutine \emph{setsub} which
is in some sense a super subroutine.

\subsection{The set of subroutines (in the order of the calls)}

\begin{enumerate}
\item The subroutine \emph{read1}: reads the basic input parameters in the
file input.dat

\item The subroutine \emph{write1}: performs some tests and writes on a
files eigvals.res and conver.res

\item The subroutine \emph{setsub}: drives the successive calculations

\item The subroutine \emph{write2}: writes on the file eigvals.res

\item The subroutine \emph{write3}: writes on the file conver.res

\item The subroutine \emph{woodsparam}: calculates the Myers parameters

\item The subroutine \emph{surfparam}: calculates the surfaces parameters

\item The subroutine \emph{basisparam}: calculates the oscillator basis
parameters

\item The subroutine \emph{pottablo} : stores the potential at the nodes of
quadrature

\item The subroutine \emph{coefftablo}: stores the products of the
coefficient of quadrature

\item The subroutine \emph{hermitablo}: stores the Hermite polynomials at
the nodes

\item The subroutine \emph{coultablo}: stores the coulomb potential at the
nodes

\item The subroutine \emph{statesplus}: selects the numbers and the
oscillator basis states corresponding to the positive parity

\item The subroutine \emph{statesminus}: selects the numbers and the
oscillator basis states corresponding to the negative parity

\item The subroutine \emph{idm}: calculates the total numbers of the used
basis states

\item The subroutine \emph{matpotplus}: calculates the representative matrix
of the central mean potential for the positive parity

\item The subroutine \emph{matpotminus}: calculates the representative
matrix of the central mean potential for the negative parity

\item The subroutine \emph{matcinplus}: calculates the representative matrix
of the kinetic energy for the positive parity

\item The subroutine \emph{matcinminus}: calculates the representative
matrix of the kinetic energy for the negative parity

\item The subroutine \emph{matpotsoplus}: calculates the representative
matrix of the mean spin-orbit energy for the positive parity

\item The subroutine \emph{matpotsominus}: calculates the representative
matrix of the mean spin-orbit energy for the negative parity

\item The subroutine \emph{diagoplus}: diagonalizes the representative
matrix of the hamiltonian for the positive parity

\item The subroutine \emph{diagominus}: diagonalizes the representative
matrix of the hamiltonian for the negative parity

\item The subroutine \emph{tred1}: Eispack subroutine (see section 8)

\item The subroutine \emph{tql1}: Eispack subroutine (see section 8)

\item The subroutine \emph{tsturm}: Eispack subroutine (see section 8)

\item The subrouine \emph{trbak1}: Eispack subroutine (see section 8)

\item The subroutine \emph{eigenvalues}: gathers the eigenvalues for both
parities

\item the subroutine \emph{vektors}: writes the eigenfunctions in a file.
\end{enumerate}

Fore several subroutines, the names ending in \textquotedblleft
plus\textquotedblright\ or in\textquotedblleft minus\textquotedblright\
means that the subroutine performs calculations specifically for a defined
parity. The term \textquotedblleft plus\textquotedblright\ is employed for
the positive parity, and the term \textquotedblleft minus\textquotedblright\
for the negative parity.

\subsection{The set of functions}

\begin{enumerate}
\item The function \emph{Hermite}: calculates the Hermite polynomials

\item The function \emph{delta}: delta symbol of Kroneker

\item The function \emph{potenv}: calculates the central mean potential
value at any point.

\item The function \emph{potenso}: calculates the spin-orbit mean potential
value at any point.

\item The function \emph{ephi}: calculates the Coulomb energy of the proton
at any point.

\item The function \emph{epslon}: estimates the round-off error for the
Eispack subroutines
\end{enumerate}

\section{Input-output data of the FORTRAN program}

If no modifications are made the use of the program as presented in long
theoretical description is very simple.

\subsection{The input data}

All input data are read from two files in a namelist type declarations. The
second file is needed only if one does use a personal parameters for the
potential, instead those of Myers. In this latter case, one has to precise
its own parameters, in a second separate file.

\subsubsection{The first input data file: \emph{input.dat}}

The file input.dat gathers all basic input data. Their significance is given
below.

\begin{itemize}
\item \emph{nmax1} and \emph{nmax2} are the bounds of the loop for $N_{\max
} $(eq(\ref{46}) or (\ref{47})). This latter is the number of the major
shells used in the calculations. If nmax1 =nmax2 (=nmax) the calculations
are performed once. The variation of nmax is envisaged only if one desires
to study the convergence of the results as a function of the number of the
basis states.

\item \emph{pi }is the pi number (3.1415927410125d.0)

\item If \emph{kkind=1}, calculations are made for the neutrons case. 
\newline
If \emph{kkind=2},calculations are made for the protons case.\newline
Any other value of the kkind parameter involves an error declaration of the
program.

\item \emph{Iz} = number of protons.

\item \emph{In} = number of neutrons.

\item \emph{Betta}, and \emph{gama} are the usual deformation parameters of
Bohr (eq.(\ref{17})-(\ref{19})).

\item If \emph{ibase=0} the states of the basis are selected according to
the spherical criterion (\ref{46}). \newline
If \emph{ibase=1}. The states of the base are selected according to the
deformed criterion (\ref{47}) There is not other value for this parameter.

\item If \emph{i1i2=1}, the program gives all eigenvalues, without
eigenvectors. \newline
If \emph{i1i2=2},the program gives the eigenvalues included in a given
interval [elow, ehigh] with the corresponding (orthonormalized)
eigenvectors. Any other value of this parameter involves an error
declaration of the program.

\item \emph{Elow} = lower bound of the selected interval.

\item \emph{Ehigh}= higher bound of the selected interval.\newline
( Naturally, if this interval is sufficiently large it will contain all
eigenvalues. Consequently all eigenvectors will be also given.)

\item If \emph{icalc=0}, the parameters of the Woods-Saxon potential are
read from the namelist of the second input file parameters dat.\newline
if \emph{icalc=1}, \ the Myers parameters are calculated by the subroutine
woodsparam.

\item If \emph{iscal=1} \ the basis parameter $\hslash \omega _{0}$ is
computed from (\ref{53}) i.e. from $\hslash \omega _{0}=0.979\left\vert
V_{0}\right\vert .A^{-1/3}$ \newline
If \emph{iscal=2} the basis parameter $\hslash \omega _{0}$ is computed from
the relation $\hslash \omega _{0}=faktor.A^{-1/3}.$(see eq.(\ref{52}).

\item \emph{faktor}= input parameter of the previous relation
\end{itemize}

\subsubsection{The second input data file: \emph{parameters.dat}}

There is an option ( governed by the keyword icalc ) in the first input file
which permits to the user to employ its own parameters instead of those of
Myers.

The data of the file parameters.dat are :

\begin{itemize}
\item \emph{v0neut}= deep of the central part of the potential for the
neutrons

\item \emph{avneut}= diffuseness of the central part of the potential for
the neutrons

\item \emph{rvneut}= radius of the central part of the potential for the
neutrons

\item \emph{capasoneu}= spin-orbit coupling strengh for the neutrons

\item \emph{assoneu}= diffuseness of the spin-orbit part of the potential
for the neutrons

\item \emph{rssoneu}= radius of the spin-orbit part of the potential for the
neutrons

\item \emph{v0pro}= deep of the central part of the potential for the protons

\item \emph{avpro}= diffuseness of the central part of the potential for the
protons

\item \emph{rvpro}= radius of the central part of the potential for the
protons

\item \emph{capasopro}= spin-orbit coupling strengh for the protons

\item \emph{assopro}= diffuseness of the spin-orbit part of the potential
for the protons

\item \emph{rssopro}= radius of the spin-orbit part of the potential for the
protons

\item \emph{rchpro}= radius of the coulomb potential \ 
\end{itemize}

\subsection{The output data}

The global results can be extracted from the five arrays \emph{evalplus,
evalminus, evecplus, evecminus,} and \emph{energies}, in the main program.%
\newline
The arrays \emph{evalplus} and \emph{evalminus} contain respectively, the
eigenvalues for the positive parity and the negative parity . The
eigenvalues are classified in an increasing order.\newline
In the same way, the arrays \emph{evecplus} and \emph{evecminus} contain the
components of the eigenvectors, in columns, in the same order as that of the
eigenvalues.\newline
For the positive parity ( respectively the negative parity) , the parameter 
\emph{nevalplus} in the subroutine diagoplus (respectively \emph{nevalminus}
in the subroutine diagominus) gives the number of eigenvalues.\newline
Sometimes, it is more convenient to gather all eigenvalues in a common array
(but the eigenvectors remain in their respective blocks). This is carried
out in a common array named \emph{energies}. In this array, the eigenvalues
are classified in an increasing order.

In order to find the corresponding eigenvector to a given eigenvalue, a
vector containing a supplemental information was created and named \emph{%
num(k)}. The sign and the absolute value of num(k) indicate respectively the
block (i.e. evecplus or evecminus) and the place of the column in this block.

Furthermore, the output data can be consulted in a straightforward way, in
three files:\newline
a) The eigenvalues are written in the file " eigvals.res ". In this file, it
is indicated in particular, if the eigenvalues belong to the set
corresponding to the positive parity or those corresponding of the negative
parity.\newline
b) The eigenfunctions are recorded in the file " vekt.res ". For every
eigenvalue there is a set of components relative to the different states
(nx, ny, nz, sigma) of the basis.\newline
c) A brief study on the convergence is made in the file " conver.res ".

\section{Checking of the computer code and comments on the test run}

In order to check the code, one has proceeded to three types of tests. In
the first, we use well-known analytical results. In the second, we compare
our calculations with those using the same model. At last, in the third, we
use some well-known properties of symmetry.

\subsection{Analytical tests}

In fact, the method of resolution of the Schrodinger equation proposed here
is a \emph{purely numerical method}. Consequently, one can use, not only the
Woods-Saxon potential, but also \emph{any other type of potential}. It is
then possible to replace the Woods-Saxon potential by that of the harmonic
oscillator in order to test the code by well-known analytical results.

\subsubsection{The deformed case without spin-orbit term}

Indeed, for a pure deformed harmonic oscillator, (without spin-orbit
interaction), in cartesian coordinates, the theoretical expression of the
energy is given simply by:\newline
$E(n_{x},n_{y},n_{z})=(n_{x}+\frac{1}{2})\hslash \omega _{x}+(n_{y}+\frac{1}{%
2})\hslash \omega _{y}+(n_{z}+\frac{1}{2})\hslash \omega _{z}$\newline
$n_{x},n_{y},n_{z}=0,1,2,.......\infty $\newline
For reasons of simplicity, we have chosen the same frequencies as those of
the basis. The numerical values are extracted from the file "eigvals1.res".%
\newline
In the computer program, one must replace the Woods-Saxon potential by that
of the harmonic oscillator, i.e. by:\newline
$V(\overrightarrow{r})=\frac{1}{2}m(\omega _{x}^{2}x^{2}+\omega
_{y}^{2}y^{2}+\omega _{z}^{2}z^{2})$\newline
Then, one has to cancel the spin-orbit interaction (by making the function
potenso = 0 or by cancelling the spin-orbit coupling constant) in the code.%
\newline
Calculating some levels analytically, and comparing them with those of the
code, one can note an excellent agreement (to seven significant digits) for
the deformed case (see table 1)\newline
It is important to note that the matrix elements are integrated numerically
in the computer code, therefore, from this test, we can conclude that the
program performs this task correctly. Because this term is diagonal ( in
fact the code "does not know this" but after calculations, it finds that the
nondiagonal elements are equal to zero ) in our basis, this test does not
permit us to verify the diagonalization. These latter part of the program
will be verified in the subsections below.\newline
Now, if we add the spin-orbit interaction, we could test the program
entirely. Unfortunately, in the deformed case, there is no theoretical
expression for that.

\subsubsection{The spherical case without, and, with spin-orbit term}

Of course, for the spherical case, it is possible to make $\hslash \omega
_{x}=\hslash \omega _{y}=\hslash \omega _{z}=\hslash \omega _{0}$ in the
previous theoretical expression. Nevertheless, the spherical coordinates are
more convenient because as we shall see, the spin-orbit term has to be
"treated" in that system. In this latter, the theoretical expression of the
energy of a pure oscillator is well-known:\newline
$E(n,l,m)=\left[ 2(n-1)+\ell +\frac{3}{2}\right] \hslash \omega _{0}=\left[
N+\frac{3}{2}\right] \hslash \omega _{0}$\newline
$N=2(n-1)+\ell =$ This number specifies a major shell\newline
$n=1,2,..........\infty $\newline
$\ell =0,1,2,.........\infty $\newline
$m=-\ell ,-\ell +1,...,\ell $, \ \ \ \ \ \ \ \ \ \ \ \ \ (for a fixed $\ell $%
)\newline
Due to the fact that the spherical symmetry is a particular case of the
deformed case, it is obvious, that the results of the code (eivals2.res )
should be in complete agreement ( with the awaited degeneracy) with the
analytical results. We can see in.table 2a,.that it is really the case.%
\newline
Furthermore, there, contrary to the deformed case, it is possible to obtain
the analytical expression for the spin-orbit term.\newline
Indeed, one can use the relation (\ref{22}) in a suitable way in order to
obtain a simple theoretical expression for the spin-orbit term.\newline
Taking $S(\overrightarrow{r})=cr^{2}/2$, where c is a positive constant, one
gets then:\newline
$-\frac{1}{r}\frac{\partial S(r)}{\partial r}\overrightarrow{\ell }.\vec{%
\sigma}=-c\overrightarrow{\ell }.\vec{\sigma}=-2c\overrightarrow{\ell }.%
\overrightarrow{s}=-2c\frac{1}{2}(\overrightarrow{j}^{2}-\overrightarrow{%
\ell }^{2}-\overrightarrow{s}^{2})$\newline
The most important point is that, in this way the splitting of the major
shells does not depend on $r$, and, is then \emph{rigorously} given by:%
\newline
$\Delta E(\ell -\frac{1}{2})=c.(\ell +1)$ \ \ \ \ \ \ \ if \ \ \ $%
j=\left\vert \ell -\frac{1}{2}\right\vert $ \ and $\ell \neq 0$\newline
$\Delta E(\ell +\frac{1}{2})=-c.\ell $ \ \ \ \ \ \ if \ \ \ \ \ $j=\ell +%
\frac{1}{2}$ \ and $\ell \neq 0$\newline
Thus, the new energies can be written as:\newline
$E(n,l,j=\ell \mp \frac{1}{2})=\left[ 2(n-1)+\ell +\frac{3}{2}\right]
\hslash \omega _{0}+\Delta E(\ell \mp \frac{1}{2})$\newline
Therefore, the code can be verified in its integrality.\newline
In order to simplify the numerical values of the splitting, we take $c=1MeV$%
. Thus, except for the value $\ell =0,$ we can see that the levels are
simply shifted by integer values according to the value of $\ell $. In order
to illustrate that, we will give two examples:

\begin{itemize}
\item Example 1 :\newline
if $\ell =1$, the $p$ shell with a energy noted $E(p)$ splits into two
subshells according to the two values of $j$:\newline
for $j=\ell +\frac{1}{2}=1+\frac{1}{2}=3/2$,\newline
$E(p3/2)=E(p)-\ell =E(p)-1Mev$ \newline
for $j=\ell -\frac{1}{2}=1-\frac{1}{2}=1/2$,\newline
$E(p1/2)=E(p)+(\ell +1)=E(p)+2Mev$

\item Example 2:\newline
similarly, if $\ell =3$ ($f$ shell ) one obtains \newline
for $j=\ell +\frac{1}{2}=3+\frac{1}{2}=7/2$, \newline
$E(f7/2)=E(f)-\ell =E(f)-3Mev$, \newline
for $j=\ell -\frac{1}{2}=3-\frac{1}{2}=5/2$, \newline
$E(f5/2)=E(f)+(\ell +1)=E(f)+4Mev$
\end{itemize}

In the table 2b, we compare all results of the code (eigvals3.res) with
those of the analytical expression. Practically, the code ( which works in
double precision) gives the exact values to six or seven significant digits
for all levels of the spectrum.\newline
This high accuracy is due to the fact that the oscillator potential is a
polynome of order two, therefore the Gauss method gives in this case the
exact values for all matrix elements.\newline
Of course, these tests are not realistic cases, but they prove that the code
runs properly with a high degree of precision.\newline
The case with spin-orbit term is very important because it involves the
integral analytical checking of the code. Due to the fact that this operator
is not diagonal in the oscillator basis, it proves not only that the code
performs correctly all calculations of the matrix elements, but also proves
that the step of the diagonalization is done properly.\newline
One also made some additional easy checks (not shown here). For example, by
taking a constant potential in the spin-orbit term , one cancels the
spin-orbit potential . That was well verified by the code, etc...

\subsection{Comparisons with similar works}

For the deformed Woods-Saxon potential , it seemed to us more convenient to
compare our code with those of the reference \cite{2}. The reasons are the
following:\newline
a) We use exactly the same model as this reference.\newline
b) All potential parameters of the calculations are precised in that
reference, and we need to use the same.\newline
c) Not only a part, but the entire spectrum of eigenvalues\ is given (as a
function of the deformation).\newline
The only disadvantage is that the results are displayed under a graphical
form. However, in extracting the numerical values, we have tried to minimize
the errors by using a graphical software.\newline
The eigenvalues are read with the own scale of the software. Then, a
suitable lin\'{e}ar transformation returns these values in MeV.
Nevertheless, in order to find the "best values", this transformation has
been carried out by the least-squares' method of the software.\newline
It turns out that it is possible to obtain values with an error about $\pm
0.03MeV.$\newline
We have thus considered the deformation $(\beta =0.3,\gamma =0.0)$ for the
lead Pb208.\newline
For the basis parameters ($Nmax$ and $\hbar \omega _{0}$), we tried to use
in calculations, the same, in order to obtain, as much as possible, close
results. For Nmax, the reference \cite{2} indicates that the matrices
corresponding to the two parities have a dimension of about 160 states.
Consequently the fixed value for Nmax was certainly Nmax=10. However, the
value of $\hbar \omega _{0}$ really used by the code is not given. This
reference indicates only that, for the spherical case, the theoretical
relation $\hbar \omega _{0}=55.A^{-1/3}$ is better than the standard
theoretical relation $\hbar \omega _{0}=41.A^{-1/3}$. Nevertheless, the
reference \cite{9}, claims that a practical value of the order of 45-48MeV
(instead $55.MeV$) gives a somewhat better results that these theoretical
relations. Since the codes of these two references have been compared, it is
probable that a common practical value was fixed. We endeavored " to guess "
this value. After many tests, It turned out that the value 47 MeV gave a
good agreement\newline
Our calculations were carried out successively with $Nmax=10$ (as the cited
reference), and $Nmax=26$. Indeed, this latter value insures that the levels
are calculated with about three or four significant digits near the fermi
level, and obviously, all the more for lower levels (see the file
"conver13.res"). They are thus practically independent of the choice of the
basis parameters\newline
In the tables (3a-3b,4a-4b) which have been deduced from the files
"eigvals4.res", "eigvals5.res", "eigvals6.res", "eigvals7.res", we show
respectively all bound levels of the Pb208 for four cases: \newline
a) \emph{neutrons-prolate shape (}$\gamma $\emph{=0}$%
{{}^\circ}%
,\beta =0.3$\emph{),Nmax=10}\newline
b) \emph{protons-prolate shape (}$\gamma $\emph{=0}$%
{{}^\circ}%
,\beta =0.3$\emph{),Nmax=10}\newline
c)\emph{neutrons-prolate shape (}$\gamma $\emph{=0}$%
{{}^\circ}%
,\beta =0.3$\emph{)Nmax=26}\newline
d)\emph{protons-prolate shape (}$\gamma $\emph{=0}$%
{{}^\circ}%
,\beta =0.3$\emph{)Nmax=26}\newline
The levels were separated in two distinct blocks according to their parity.%
\newline
Of course, for a finite potential, the discrete positive energy levels do
not represent, a valid solution of the continuum (see ref.\cite{10}),
therefore, we shall drop them .\newline
In all cases, we can note that the energy levels are practically the same
ones for the low part of the spectrum, but relative small differences appear
in the \emph{upper part of the spectrum}.

These differences are more prounonced for $Nmax=26$ that for $Nmax=10$. The
analyse of these results leads to the following conclusions:

\begin{itemize}
\item The lowest levels of the spectrum converge systematically more quickly
than the others. As one goes up in the spectrum the convergence is in
general slowest, but there can be some rare exceptions.

\item A rapid convergence involves a weak dependence relatively to the basis
parameters. The highest levels of the spectrum are thus more sensitive to
the basis parameters. One can affirm that if the basis parameters of our
code are close to those of the reference \cite{2}, they are not rigorously
the same ones.

\item In fact, one noted that this remark is general. Indeed, a modification
of any parameter (for example those of the potential) in the calculations
produces a modification relatively more significant for the highest levels
than for the lowest levels. For example if the radius of the mean potential
(spherical case for the neutrons) varies from 7.36fm to 7.40fm (all other
parameters being constant), the first level, and the Fermi level undergo
variations of 0.05MeV, and 0.28MeV respectively. The "general rule" is thus
that \emph{the lowest levels are most "stable"}.

\item Owing to the fact that we employ very similar parameters, our results
with $Nmax=10$ are "artificially" very close to those of the reference \cite%
{2}(the mean deviations are about 0.05 MeV for all cases).\ Thus, our
purpose which was to recover the same results is now reached. But, the word
"artificially" means that for this small basis, both results are not enough
accurate, although they are the same \newline
Indeed, it is clear that they will be actually less precise than those
obtained with $Nmax=26$. Significant differences appear in the top of the
spectrum. In the file "conver13.res", one can note that the Fermi level is
stabilized to about $0.01\backsim 0.03$ $MeV$ only starting from $%
Nmax=15\backsim 16$. Therefore, calculations with $Nmax=10\backsim 14$
produce mediocre results.
\end{itemize}

The rapid convergence of the lowest states is due mainly to the fact that
the corresponding wave functions are very similar to those of the
oscillator. This is not the case for the highest states where the wave
functions are strongly oscillating, and where the edge effect of the
potential is "felt".\newline
This can be easily noticed in the components of the eigenvectors, in the
file "vekt14.res" . For example, concerning the first eigenvalue, only the
components corresponding to the lowest quantum numbers are important (see
the components numbered 1, 2, 12, and, 59).

\subsection{Tests using some properties of the parametrisation $(\protect%
\beta ,\protect\gamma )$}

Two simple tests can be carried out to check the consistence of the program:

In the first, one compares the spectra obtained with the deformations$(\beta
,\gamma )$ and $(\beta ,-\gamma )$ This operation is in fact nothing other
that a simple permutation of the axes 1 and 2 of the ellipsoid. Of course
the two shapes are the same, consequently, the respective spectra must be
identical.

In the files "eigvals8.res" and "eigvals9.res" one can easily check that is
really the case with an astonishing precision. In particular, one can note
in these files the permutation of values of the parameters $\hbar \omega
_{x} $(hbaromegx), and $\hbar \omega _{y}$(hbaromegy).

In the second, one compares the spectra obtained with the deformations $%
(\beta ,\gamma =60%
{{}^\circ}%
)$ and $(-\beta ,\gamma =0%
{{}^\circ}%
)$. There also, this operation is simply a cyclic permutation of the three
axes of the ellipsoid. Therefore, the spectrum must also remain unchanged.

As for the previous case, this can be easily verified in the files
"eigvals10.res and eigvals11.res".

\subsection{Tests of convergence}

In the files "conver12.res" and "conver13.res", we have shown the
convergence of the sum (of the single particle energy) of the first 126
neutrons levels of Pb208 for two deformations. The potential's parameters
are those of the reference \cite{2}.

This sum has converged to less than 1 Mev only starting from the values $%
Nmax=14$ and $Nmax=16$, respectively for the spherical and the deformed
cases.\newline
This implies for the Fermi level, a convergence to 0.02 MEV and 0.01 MEV
respectively for these two cases. However, for $Nmax\backsim 16$, theses
deviations depend still of the value $\hbar \omega _{0}$. Obviously, for
higher bound states , the precision will be less.\newline
Everything depends on what one wants to make. So, for example , for the
Strutinsky's shells corrrection the previous values seem to be sufficient.

Always concerning the Fermi's level (conver13.res), one notices in general
that it increases in absolute value as $Nmax$ increases, but sometimes, it
happens that it decreases slightly (in absolute value). For example, in the
spherical case, it passes from 8.510 to 8.502 when Nmax passes from 13 to
14. We can easily see that the dimensions of subspaces corresponding to the
positive and negative parity do not vary simultaneously when Nmax varies by
one unit. For example, when one passes from Nmax=13 to Nmax=14 only one
subspace, namely the one with a positive parity, undergoes changes from 252
basis states to 372 basis states. The other remains the same with 308 basis
states.\newline
In our example, the Fermi level belongs to the subspace of negative parity,
therefore, apparently, it should not have to change. In fact, the formulae
of the spin-orbit interaction connects the matrix elements of the two
subspaces (see eq.\ref{38}-\ref{40}). This implies always a slight
modifications in the subspace which has not varied, and this must not be
assimilated to a noise.

In fact, in this method, the "true noise" has two main sources :

a) under-estimations of the number of points in the numerical integrations
of the matrix elements.

b) a too small basis or really inadequate values of the basis parameters.

With 30 points of quadrature, a double precision, and a large basis ($Nmax$ $%
\ $up to 26) these two problems are here minimized.

\section{Conclusion}

We have elaborated and checked a calculation program solving the equation of
Schrodinger for a deformed potential of Woods-Saxon type.\newline
The program appears very rapid, and consequently, it becomes possible to use
significant basis sizes.\newline
Calculations with small bases, like those which were carried out in the past
with $Nmax=10\backsim 12$ lead to a very poor precision. Our conclusions are
corroborated by other works. For example, the ref.\cite{5} has shown for
Hartree-Fock calculations that the error in the energy of Pb208 is smaller
than 1MeV only for $Nmax\geqq 16$. Other examples are given in the ref.\cite%
{4} which confirm this fact.\newline
Similar codes \cite{6,7} were made in the past, but with the assumption of
axial symmetry. To our knowledge, triaxial Woods-Saxon calculations were
never really undertaken with significant sizes of the oscillator basis.

\appendix

\section{The Myers parameters}

The diffuseness parameters $a_{V}$, $a_{so}$, and the spin-orbit coupling $%
\kappa $ are the same as those of the Ref.\cite{1}.%
\begin{equation}
a_{V}=0.66\ fm
\end{equation}%
\begin{equation}
a_{so}=0.55\ fm
\end{equation}%
\begin{equation}
\kappa =12\ MeV\ fm^{2}
\end{equation}%
The parameters of central potential, and of the spin-orbit potential, were
extracted from the Myers droplet model \cite{3,8}. This theory uses Thomas
Fermi's approximation to approach average properties of finite nuclei like
the density radii, skin-thicknesses, ..., in terms of neutron and proton
numbers.\newline
In this model, two auxiliary quantities are first defined:%
\begin{equation}
\widehat{\delta }=\frac{\frac{N-Z}{A}+0.0112\frac{Z^{2}}{A^{5/3}}}{1+\frac{%
3.15}{A^{1/3}}}
\end{equation}%
\begin{equation}
\widehat{\epsilon }=-\frac{0.147}{A^{1/3}}+0.330\widehat{\delta }^{2}+\frac{%
0.00248Z^{2}}{A^{4/3}}
\end{equation}%
The physical significance of these quantities is explained in the Ref.\cite%
{3} \newline
With help of these quantities, the depth of the mean potentials are written
as: 
\begin{equation}
V_{0}(protons)=-52.5-48.7\widehat{\delta }
\end{equation}%
\begin{equation}
V_{0}(neutrons)=-52.5+48.7\widehat{\delta }
\end{equation}%
The radii of the central potentials (which are different for protons and
neutrons) are expressed by means of the nuclear density radii $%
R_{0}(protons) $, or, $R_{0}(neutrons)$, and the density diffuseness $a_{V}$:%
\begin{equation}
R_{V}(protons)=R_{0}(protons)\left\{ 1-\frac{\pi ^{2}}{3}\left( \frac{a_{V}}{%
R_{0}(protons)}\right) ^{2}\right\}
\end{equation}%
\begin{equation}
R_{V}(neutrons)=R_{0}(neutrons)\left\{ 1-\frac{\pi ^{2}}{3}\left( \frac{a_{V}%
}{R_{0}(neutrons)}\right) ^{2}\right\}
\end{equation}%
with\newline
\begin{equation}
R_{0}(protons)=R_{0}+0.82-\frac{0.56}{R_{0}}+0.22\widehat{\delta }
\end{equation}%
\begin{equation}
R_{0}(neutrons)=R_{0}+0.82-\frac{0.56}{R_{0}}-0.22\widehat{\delta }
\end{equation}%
\begin{equation}
R_{0}=r_{0}A^{1/3}(1-\widehat{\epsilon })
\end{equation}%
\begin{equation}
r_{0}=1.16\ \ fm
\end{equation}%
The radius of the spin-orbits mean field is given in the same way:%
\begin{equation}
R_{so}=R_{0}\left( 1-\frac{\pi ^{2}}{3}\left( \frac{a_{so}}{R_{0}}\right)
^{2}\right)
\end{equation}%
At last, the radius of the charge density is given by:\newline
\begin{equation}
R_{ch}=R_{0}-\frac{1}{3}r_{0}A^{1/3}\left( \frac{N-Z}{N+Z}-\widehat{\delta }%
\right)
\end{equation}


\begin{thebibliography}{10}
\bibitem[1]{1} H.C.Pauli, Physics Report (Phys.Lett.C) 7 (1973) 35

\bibitem[2]{2} U.Gotz, H.C.Pauli, K.Alder, Nucl.Phys. A175 (1971) 481

\bibitem[3]{3} W.D. Myers, Nucl.Phys. A145 (1970)387

\bibitem[4]{4} T.Vertse, A.T.Kruppa, W. Nazarewicz , Phys.Rev.
C61(2000)064317

\bibitem[5]{5} J.Dobaczewski,J. Dudek, Comput.Phys.Commun. 102 (1997) 183

\bibitem[6]{6} Garrote, R. Capote, R. Pedrosa. Comput. Phys. Commun.
92(1995)267

\bibitem[7]{7} S. Cwiok, J. Dudek, W. Nazarewicz, J. Skalski, T.R. Werner.
Comput. Phys. Commun. 46(1987)379

\bibitem[8]{8} P.Moller,J.R.Nix,W.D.Myers,W.J.Swiatecki. Atom. Data Nucl.
Data Tabl.59(1995)185-381

\bibitem[9]{9} J. Damgaard, H.C. Pauli, V.V. Pashkevich, V.M. Strutinsky,
Nucl. Phys. A135(1969)432

\bibitem[10]{10} M. Bolsterli, E.O. Fiset, J.R. Nix, and J.L. Norton,
Phys.Rev. C5(1972)1050
\end{thebibliography}
\end{document}